\documentclass[preprint,12pt]{elsarticle}

\usepackage{amssymb}
\usepackage{amsmath}
\usepackage{graphicx} 
\usepackage{amsfonts}
\usepackage{natbib}
\usepackage{textcomp}
\usepackage{float}
\usepackage{subcaption}
\usepackage{textcomp}
\usepackage{acronym}
\usepackage{pdflscape}
\usepackage{comment}
\usepackage{array}
\usepackage{booktabs}
\usepackage[hidelinks]{hyperref}
\newcolumntype{P}[1]{>{\centering\arraybackslash}p{#1}}
\newcolumntype{M}[1]{>{\centering\arraybackslash}m{#1}}
\usepackage{etoolbox}
\usepackage{cleveref}
\usepackage{eurosym}
\usepackage{indentfirst}
\usepackage{adjustbox}
\usepackage{nicefrac}
\usepackage{multirow}
\usepackage{subcaption}
\usepackage{bbding}
\usepackage[table]{xcolor}
\usepackage{pifont}
\newcommand{\cmark}{\ding{51}}
\newcommand{\xmark}{\ding{55}}
\usepackage{glossaries}
\usepackage{makecell}

\newacronym{coi}{COI}{center of inertia}
\newacronym{ufls}{UFLS}{under-frequency load shedding}
\newacronym{rocof}{RoCoF}{rate of change of frequency}
\newacronym{uc}{UC}{unit commitment}
\newacronym{milp}{MILP}{mixed integer linear programming}
\newacronym{ips}{IPS}{island power system}
\newacronym{res}{RES}{renewable energy source}
\newacronym{ed}{ED}{economic dispatch}
\newacronym{fcuc}{FCUC}{frequency constrained unit commitment}
\newacronym{sfr}{SFR}{system frequency response}

\journal{International Journal of Electrical Power \& Energy Systems}

\usepackage[intoc]{nomencl}
\usepackage{xstring}
\usepackage{xpatch}
\usepackage{eurosym}

\patchcmd{\thenomenclature}
  {\leftmargin\labelwidth}
  {\leftmargin\labelwidth\itemindent 1em }
  {}{}
\newcommand{\nomenclheader}[1]{%
  \item[\hspace*{-\itemindent}\normalfont\bfseries#1]}
\renewcommand\nomgroup[1]{%
  \IfStrEqCase{#1}{%
   {A}{\nomenclheader{Indices and sets}}
   {B}{\nomenclheader{Parameters}}
   {C}{\nomenclheader{Variables}}
  }%
}

\makenomenclature

\nomenclature[A]{\(i\)}{Index of generators}
\nomenclature[A]{\(t\)}{Index of time intervals}
\nomenclature[A]{\(\ell\)}{Index of the lost unit}
\nomenclature[A]{\(s\)}{Alias index for time intervals}
\nomenclature[A]{\(\ell e\)}{Index of the leaves}
\nomenclature[A]{\(\mathcal{T}\)}{Set of all time intervals}

\nomenclature[B]{\(D\)}{Load damping factor}
\nomenclature[B]{\(H\)}{Inertia [s]}

\nomenclature[C]{$\frac{d\Delta f(t)}{dt}$}{Initial \gls{rocof} [Hz/s]}

\begin{document}

\begin{frontmatter}



\title{A Corrective Frequency-Constrained Unit Commitment with Data-driven Estimation of Optimal UFLS in Island Power Systems}


\author[inst1]{Miad Sarvarizadeh}
\author[inst1]{Lukas Sigrist}
\author[inst2]{Almudena Rouco}
\author[inst1]{Mohammad Rajabdorri}
\author[inst1]{Enrique Lobato}

\affiliation[inst1]{organization={IIT, Comillas Pontifical University},
            city={Madrid},
            country={Spain}}
\affiliation[inst2]{organization={ICAI, Comillas Pontifical University}, 
city={Madrid},
country={Spain}}

\begin{abstract}
This paper presents a novel corrective \gls{fcuc} formulation for island power systems by implementing data-driven constraint learning to estimate the optimal \gls{ufls}. The Tobit model is presented to estimate the optimal amount of \gls{ufls} using the initial rate of change of frequency. The proposed formulation enables co-optimizing operation costs and \gls{ufls}. The aim is to account for optimal \gls{ufls} occurrences during operation planning, without increasing them. This would potentially reduce system operation costs by relaxing the reserve requirement constraint. The performance of the proposed formulation has been analyzed for a Spanish island power system through various simulations. Different daily demand profiles are analyzed to demonstrate the effectiveness of the proposed formulation. Additionally, a sensitivity analysis is conducted to demonstrate the effects of changing the cost associated with \gls{ufls}. The corrective \gls{fcuc} is shown to be capable of reducing system operation costs without jeopardizing the quality of the frequency response in terms of \gls{ufls} occurrence.
\end{abstract}

\glsresetall
\begin{highlights}
\item A novel corrective frequency-constrained unit commitment formulation is presented that considers and co-optimizes optimal under-frequency load-shedding amounts with system costs such that the system is secure after the outage of each generation unit, and operation costs are reduced by relaxing the reserve requirement.
\item A data-driven constraint learning approach is proposed and implemented to estimate the amount of optimal under-frequency load shedding using the Tobit model.
\item  The estimating model is linearized and presented in mixed integer linear programming form to facilitate its addition to the standard unit commitment problem.
\item The proposed formulation is applied to a real Spanish island power system. The effectiveness and capability of the formulation are analyzed through extensive simulations and sensitivity analyses.
\end{highlights}

\begin{keyword}
frequency-constrained unit commitment \sep data-driven \sep Tobit model \sep frequency stability \sep machine learning \sep corrective

\end{keyword}

\end{frontmatter}


\section{Introduction}\label{sec:intro} 
The high penetration of \glspl {res} in power systems has resulted in reduced inertia, as \glspl{res} do not inherently provide inertia. This issue is particularly critical in \glspl{ips}, where inertia is already limited due to the small number of available generating units \cite{wang2013high}. The reduced inertia in \glspl{ips} lowers the overall frequency control capability, making the system more vulnerable to rapid frequency declines following the loss of a single generating unit. It is critical to address this issue in the scheduling problem of \glspl{ips} to avoid large frequency deviations.

In most \glspl{ips}, the scheduling problem is centrally managed. The system operator sequentially performs \gls{uc} and \gls{ed} over different time horizons. Optimal scheduling is typically constrained by security of supply requirements. A common security constraint is ensuring a sufficient amount of spinning reserve, which is activated in response to disturbances that cause an active power imbalance.
Currently, the spinning reserve requirement follows a static criterion such that the amount of spinning reserve is sufficient to cover the largest possible contingency in the \glspl{ips}, following an N-1 security criterion.

A known limitation of the static spinning reserve criterion is that it does not necessarily guarantee its timely activation \cite{thalassinakis2007method}. After a disturbance, the spinning reserve is activated in real-time through primary and secondary frequency controls, where the primary frequency control restricts frequency variations and the secondary frequency control brings frequency back to its nominal value. However, in low-inertia systems, primary frequency control cannot always fully activate the spinning reserve in time to effectively contain frequency deviations. As a result, \gls{ufls} is triggered to arrest frequency decay even under moderate disturbances in \glspl{ips}. This is mainly because the turbine-governor systems inherently exhibit a technology-dependent delay in their dynamic response to increase power. 

The dynamic nature of the problem of spinning reserve activation within the scheduling problem has already been addressed in the literature under the framework of \gls{fcuc}. This includes adding limitations on the post-contingency frequency characteristics such as \gls{rocof}, frequency nadir, and quasi-steady-state frequency \cite{chavez2014governor,ahmadi2014security,farrokhabadi2016unit,ferrandon2022inclusion,badesa2019simultaneous,li2023frequency}. The frequency nadir was calculated in \cite{chavez2014governor} under the assumption of a constant ramp rate in the overall mechanical power response, which is then used to formulate the corresponding constraints. The piecewise linearization technique is used to linearize the nonlinear function of frequency nadir in \cite{ahmadi2014security}, forming a \gls{fcuc} formulation. The frequency response model for a multi-machine system is used to achieve higher accuracy. A mathematical model was proposed in \cite{farrokhabadi2016unit} to represent changes in dispatchable units, assuming a linear transition rather than the typical staircase profile used in standard \gls{uc}. The reserve constraint is then modified to ensure the feasibility of frequency regulation in the system. Separable programming is used in \cite{ferrandon2022inclusion} to linearize and add the frequency nadir constraint to the \gls{uc} problem. In \cite{badesa2019simultaneous}, the speed of response from different sources is modeled through distinct ramp rates. The resulting frequency nadir constraint is then linearized and added to the \gls{uc} formulation. Similar to the previous reference, \cite{li2023frequency} proposes constraints for frequency nadir, \gls{rocof}, and quasi-steady-state frequency addressing diverse frequency support resources and then linearizes the frequency nadir constraint using the linearized-frequency method.

Data-driven constraint learning is proposed as an alternative to the analytical modeling of constraints. Constraint learning is the process of learning a function using data that relates the outcomes and decisions, to represent a constraint that is not clearly known or is highly nonlinear \cite{maragno2023mixed}. The learned function can then be used to generate predictions for a new observation. This method enhances the process of data-driven decision-making by combining machine learning techniques with optimization. Constraint learning has been used in various problems in the power systems field. In \cite{cremer2018data}, an ensemble learning method is used to describe the security boundary of the system. A security-constrained AC optimal power flow is presented in \cite{halilbavsic2018data} that ensures small-signal stability and N-1 security, utilizing decision trees. \cite{gutierrez2010neural} proposes a neural network method to obtain stability and security constraints in optimal power flow problems.

While constraint learning has been applied to various power system studies, its role in frequency stability modeling is particularly relevant \cite{rajabdorri2023inclusion,lagos2021data,sang2023conservative,liu2023modeling}. Logistic regression and support vector machines were proposed in \cite{rajabdorri2023inclusion} to predict the frequency nadir, with the resulting linear constraints incorporated into the \gls{uc} problem to represent these models. An optimal classifier tree is proposed in \cite{lagos2021data} to form a \gls{fcuc} that prevents frequency violations corresponding to the activation thresholds of the first \gls{ufls} relays. In \cite{sang2023conservative}, a conservative sparse neural network method is given that approximates frequency nadir constraints and represents them in \gls{milp} form. An extreme learning machine-based method is utilized in \cite{liu2023modeling} to approximate the nonlinear frequency nadir constraint by a set of linear constraints.

The above-mentioned literature is entirely focused on preventive measures to restrict different post-contingency frequency performance features. Apart from preventive formulations, \gls{fcuc} can consider corrective actions, including potential \gls{ufls}. The inclusion of corrective actions typically results in lower system operation costs compared to relying solely on preventive measures.
In the case of small \glspl{ips}, \gls{ufls} is unavoidable in the event of large disturbances \cite{concordia1995load, moya2005spinning}. Moreover, common preventive constraints to avoid \gls{ufls} entirely, would lead to infeasibility in the scheduling problem \cite{rajabdorri2022robust}. In this context, having large amounts of spinning reserves available might be inefficient.

 A first step towards a corrective \gls{fcuc} has been presented in \cite{teng2017full}, where an analytical expression has been proposed to determine \gls{ufls} by assuming that total generation increases linearly in time and that the disturbance is known. The resulting non-linear expression has been linearized by assuming given discrete \gls{ufls} blocks, leading to a set of bilinear constraints, which are linearized using K-block piecewise linear functions. Then in \cite{o2021probabilistic} the previous work is extended by differentiating between fast and slow generation, leading to a nonconvex expression to estimate \gls{ufls}. Similarly, discrete \gls{ufls} blocks are assumed, which allows defining a set of convex second-order cone constraints that approximate the nonconvex expression. It is observed that these previous studies on corrective \gls{fcuc} assume that overall generation responses increase linearly in time and within a fixed time. This assumption follows minimum grid code requirements, but actual responses are not linear in time, and individual responses may differ significantly according to the generation technology. Additionally, \gls{ufls} blocks have been assumed to make the corrective \gls{fcuc} convex, but such blocks are usually activated at different frequency levels, which is fully neglected. Finally, either the disturbance size is known or just the outage of the largest generation unit is considered. However, in \glspl{ips} even the outage of medium-sized generation units can lead to \gls{ufls} \cite{padron2015reducing}.  

 Few previous studies have been conducted that attempt modeling \gls{ufls} in the scheduling problem of \glspl{ips}.
 An analytical method to include \gls{ufls} as a corrective measure to the \gls{uc} problem, forming a corrective \gls{fcuc} formulation is proposed in \cite{rajabdorri2024unit}. In \cite{rajabdorri2025data} a regression tree structure is proposed to estimate the amount to \gls{ufls} based on conventional and common \gls{ufls} schemes. \Cref{tab:ref} presents a comparison of the related work in the literature with this paper.
 \begin{table}[t!]
    \centering
    \caption{Comparison of the most related papers}
    \begin{adjustbox}{width=1\textwidth}
    \begin{tabular}{c cccc}
    \toprule
        Paper & Formulation & Modeling method & \gls{ufls} scheme & \gls{ips} \\ \midrule
        \cite{chavez2014governor} & preventive & analytical & \xmark & \xmark \\
        \cite{ahmadi2014security} & preventive & analytical & \xmark & \xmark \\
        \cite{farrokhabadi2016unit} & preventive & analytical & \xmark & \xmark \\
        \cite{ferrandon2022inclusion} & preventive & analytical & \xmark & \cmark \\
        \cite{badesa2019simultaneous} & preventive & analytical & \xmark & \xmark \\
        \cite{li2023frequency} & preventive & analytical & \xmark & \xmark \\
        \cite{rajabdorri2023inclusion} & preventive & data-driven & \xmark & \cmark \\
        \cite{lagos2021data} & preventive & data-driven & \xmark & \cmark \\
        \cite{sang2023conservative} & preventive & data-driven & \xmark & \xmark \\
        \cite{liu2023modeling} & preventive & data-driven & \xmark & \xmark \\
        \cite{teng2017full} & corrective & analytical & adaptive & \xmark \\
        \cite{o2021probabilistic} & corrective & analytical & adaptive & \xmark \\
        \cite{rajabdorri2024unit} & corrective & analytical & adaptive & \cmark \\
        \cite{rajabdorri2025data} & corrective & data-driven & conventional & \cmark \\
        this paper & corrective & data-driven & conventional/adaptive & \cmark \\ \bottomrule
    \end{tabular}
    \end{adjustbox}
    \label{tab:ref}
\end{table}

This paper presents a novel corrective \gls{fcuc} formulation that incorporates the optimal \gls{ufls} amounts in the scheduling problem of \glspl{ips}.
First, a data-driven model is proposed to estimate the amount of optimal \gls{ufls} that prevents the frequency from falling below a certain threshold for a specified period, after a disturbance. Correlations between the amount of optimal \gls{ufls} and several explanatory variables have shown that the amount of optimal \gls{ufls} can be estimated using regression models \cite{sigrist2014ufls}. Several variables can be used as the feature in the model (demand, reserve, etc.) \cite{xiao2019deep}. However, the initial \gls{rocof} is shown to be highly correlated with the optimal amount of \gls{ufls} in \glspl{ips}. Then, this estimation is incorporated in the reserve constraint of the proposed corrective \gls{fcuc} formulation. The aim is thus to account for \gls{ufls} during operation planning, which could reduce system operation costs by reducing the spinning reserve accordingly. Note that, in this process, an increase of \gls{ufls} is not sought, but rather the unavoidable and optimal \gls{ufls} amounts are accounted for.
 
The main contributions of this paper are as follows:

\begin{itemize}
    \item A novel corrective \gls{fcuc} formulation is presented that considers and co-optimizes optimal \gls{ufls} with system costs such that the system is secure after the outage of each generation unit, and operation costs are reduced by relaxing the reserve requirement.
    \item A data-driven constraint learning approach is proposed and implemented to estimate the amount of optimal \gls{ufls} using the Tobit model, avoiding the need to make assumptions on the generation responses and the \gls{ufls} blocks.
    \item The estimating model is linearized and presented in \gls{milp} form to facilitate its addition to the standard \gls{uc} problem and form a corrective \gls{fcuc} formulation.
    \item The proposed formulation is applied to a real Spanish \gls{ips}. The effectiveness and capability of the formulation are analyzed through extensive simulations and sensitivity analyses.
\end{itemize}

The rest of this paper is organized as follows: \Cref{sec:datadriven} provides the data-driven model that estimates optimal \gls{ufls} amounts. Then, the corrective \gls{fcuc} formulation is presented in \Cref{moduc}. \Cref{sec:result} analyzes the performance of the proposed formulation through various simulations and sensitivity analyses. Finally, \Cref{sec:conclusion} gives the conclusions of this study. 

\section{Data-driven estimation of optimal UFLS}\label{sec:datadriven}
\subsection{Background}\label{Background}
Following the outage of a generator that results in an active-power imbalance, the frequency dynamics of an \gls{ips} can be described by the swing equation as follows, assuming the \gls{coi} model is applied:
\begin{equation}
    2H \frac{d\Delta f(t)}{dt} + D \Delta f(t)  = r(t) - p_{\ell} \label{swing}
\end{equation}
where $H$ is the equivalent system inertia, $D$ is the load-damping factor, $\Delta f(t)$ is the frequency deviation, $p_{\ell}$ is the amount of power imbalance, and $r(t)$ is the primary frequency response provided by the synchronous generators collectively.
Suppose the overall generation response was modeled by a linear first-order model with a time constant $T$ and equivalent gain of the turbine-governor system, $K$. In that case, the frequency response can be obtained as follows \cite{anderson1990low}:
\begin{equation}
    \Delta f(t)  = \frac{-p_{\ell}}{D+K}\big(1+a e^{-\zeta\omega_{n} t} sin(\omega_{r} t + \phi)\big) \label{freqresponse}
\end{equation}
where $a$, $\omega_n$, $\omega_r$, $\zeta$, and $\phi$ are parameters of the frequency response that depend on $H$, $T$, and $K$.
\cref{freqresponse} shows that the frequency exhibits an oscillatory response being damped over time. The damping depends on the undamped natural frequency, $\omega_n$, and the damping ratio, $\zeta$. The oscillation frequency is governed by the damped frequency, $\omega_r$. In steady-state, the post-disturbance frequency deviation depends on the disturbance, $K$, and $D$.

The instant of nadir frequency, $t_{\text{\tiny nadir}}$, can be computed by setting the derivative of \cref{freqresponse} to 0:
\begin{equation}
    t_{\text{\tiny nadir}}  = \frac{1}{\omega_{r}} \tan^{-1}\left(\frac{\omega_{r} T}{\zeta\omega_{n} T - 1}\right) \label{nadirtime}
\end{equation}

Using \cref{freqresponse} and \cref{nadirtime}, the critical active-power disturbance, $p_{\text{\tiny crit}}$, can be computed that would result in a frequency nadir deviation that is still acceptable according to grid codes. Note that this computation is highly nonlinear. Now, a disturbance $p_{\text{\tiny dist}}$ larger than $p_{\text{\tiny crit}}$ would at least require an amount of optimal \gls{ufls}, $p^{\text{\tiny UFLS}}$:
\begin{equation}
        p^{\text{\tiny UFLS}}  = p_{\ell} - p_{\text{\tiny crit}} = -2H\Delta \dot f_{\text{\tiny initial}} - p_{\text{\tiny crit}} \label{ufls}
\end{equation}
where $\Delta \dot f_{\text{\tiny initial}}$ is the initial \gls{rocof}. The term optimal is used to emphasize that the amount of \gls{ufls} is calculated based on the difference between the disturbance and $p_{\text{\tiny crit}}$. This amount is the lowest possible \gls{ufls} to prevent frequency from falling below the threshold. However, the conventional \gls{ufls} scheme used in the operation of \glspl{ips} works in a step-wise manner. Therefore, it might shed more load than necessary and is sub-optimal.


It is observed that the system responses in terms of frequency and specifically \gls{ufls} depend on the equivalent inertia and the magnitude of the disturbance, which are both represented within the initial \gls{rocof}.
The following subsections discuss the relationship observed between the initial \gls{rocof} and the optimal \gls{ufls} and present a data-driven methodology that clarifies their relationship. 

\subsection{Dataset generation}\label{dataset}

To construct a dataset that can represent the performance of a system, a synthetic data generation method is used similar to \cite{rajabdorri2023inclusion}.
The dataset in this paper is generated using the data of the case study, which is a Spanish \gls{ips}. Details and characteristics of this system are further presented in \Cref{case}.
To start off the dataset generation process, a comprehensive dataset containing all the possible generation point combinations that are feasible regarding the \gls{uc} constraints is produced with a desired step size. Note that the step size can be selected to obtain the required accuracy in exchange for more computational time. Then, the cheaper combinations that are closer to the solution of the optimization problem are picked out.

The next step is to label the data with the amount of optimal \gls{ufls} that would be needed after the outage of each generating unit for every remaining generation combination. This could be done by using \cref{ufls} together with \cref{freqresponse} and \cref{nadirtime}, assuming a single and linear generator model. Alternatively, the optimal amount of \gls{ufls} can be computed from the dynamic system response, simulated by using either fully detailed power system simulators or by \gls{sfr} models, which are models able to reflect short-term frequency dynamics. In this case, a \gls{sfr} is used to determine the optimal amount of \gls{ufls} for each outage such that the frequency does not fall below 48 Hz for more than 2 s and below 47 Hz (e.g., see \cite{sigrist2011method}).

\begin{figure}[t!]
    \centering
    \includegraphics[width=1\linewidth]{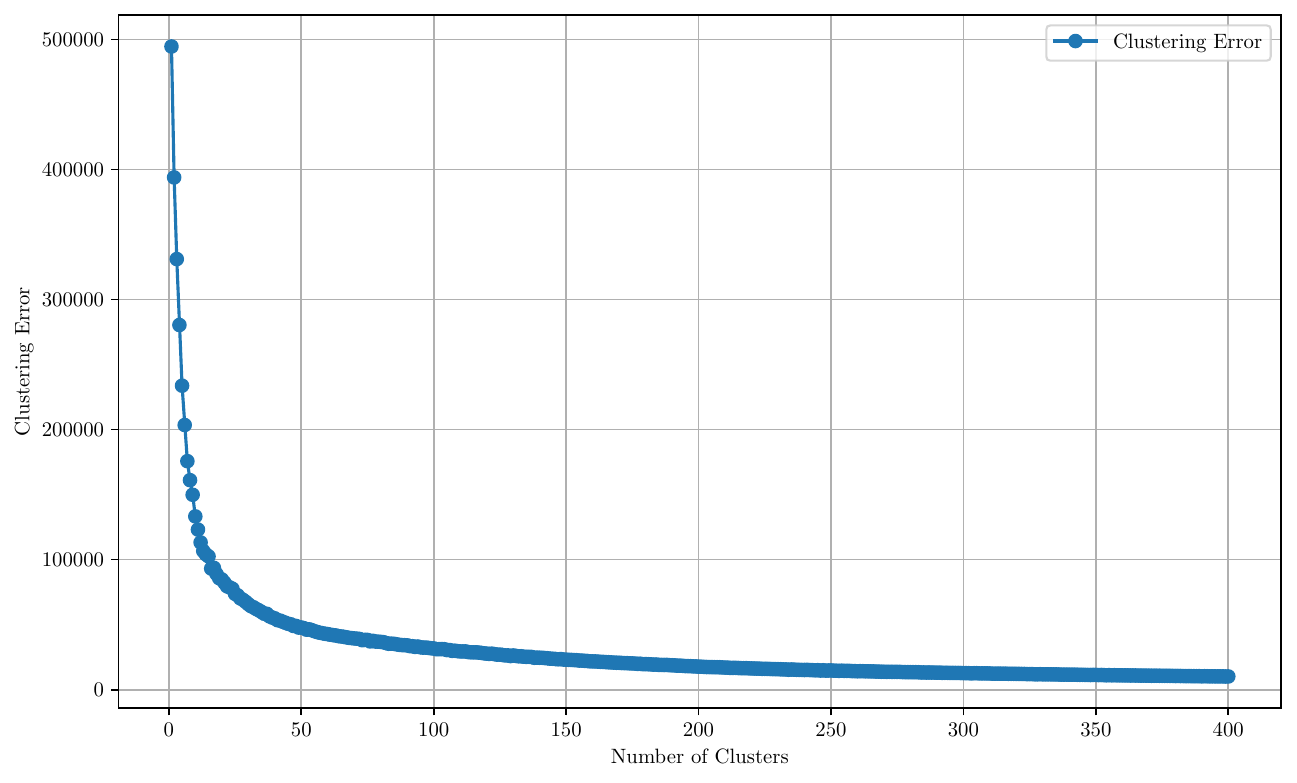}
    \caption{Clustering error vs. number of clusters}
    \label{fig:cluster}
\end{figure}

This process could be impractical in terms of computational time. For instance, for the case study of the real Spanish island used in this paper, the dataset includes 133717 points. To address this problem, a k-means clustering algorithm is used to reduce the number of data points while preserving the dataset’s representativeness.
The number of clusters was chosen so that further increasing it would not improve the clustering error by more than 0.1\%. 
\Cref{fig:cluster} shows the clustering error computed when varying the number of clusters up to 400.

\begin{figure}[t!]
    \centering
    \includegraphics[width=1\linewidth]{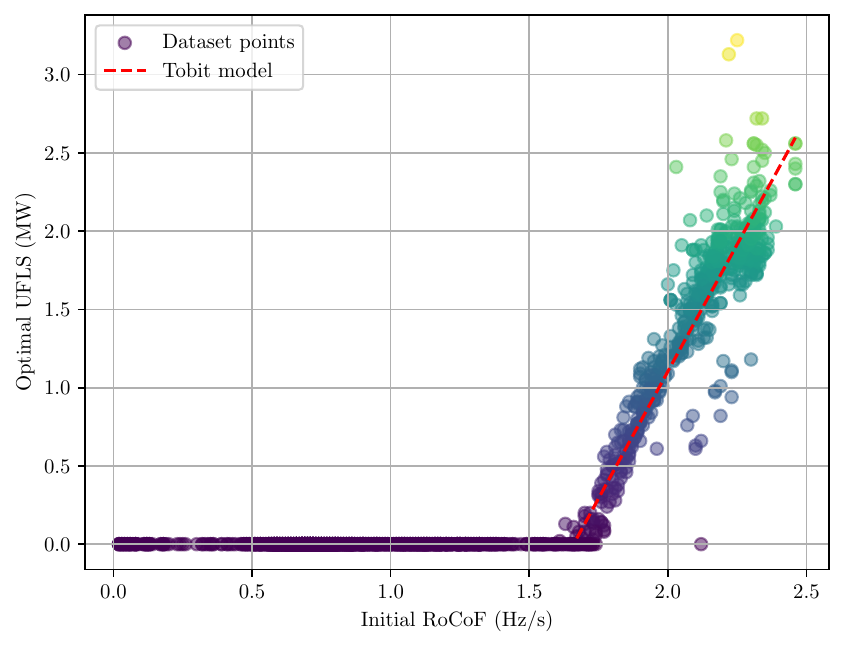}
    \caption{The points in the clustered dataset and the resulting Tobit model}
    \label{fig:dataset}
\end{figure}

The clustering error here represents the compactness of clusters by calculating the sum of squared distances between each data point and its assigned cluster centroid. The resulting dataset consists of 273 centroids. These centroids were then labeled with the corresponding optimal \gls{ufls} amount for the outage of every possible unit. \Cref{fig:dataset} shows the correlation between the initial \gls{rocof} and the resulting optimal \gls{ufls} for possible outages in all of these operating points. Based on the correlation observed in \Cref{fig:dataset},  it is evident that the amount of optimal \gls{ufls} can be estimated using the initial \gls{rocof} through functions learned from the dataset. 

\subsection{The Tobit model and its MILP representation}\label{model}
The function that relates the initial \gls{rocof} to the optimal \gls{ufls} must be \gls{milp} representable to be embedded into the \gls{uc} problem. This may require multiple constraints and auxiliary variables. Here, the Tobit model that is suitable according to the characteristics of the dataset is presented with the resulting constraints.

The Tobit model is a regression framework designed to handle censored data by accounting for observations that fall below or above a certain threshold \cite{tobin1958estimation}. Originally developed in econometrics, it is employed as a maximum-likelihood estimator for probability density function parameters in cases where data are partially observed due to truncation or censoring \cite{hampshire1992tobit}. This model is particularly useful in power system applications where certain conditions lead to natural truncation. In the dataset presented in \Cref{fig:dataset}, the optimal \gls{ufls} are inherently non-negative and exhibit a near-linear relationship above zero, making the Tobit model an appropriate choice for accurately capturing the underlying frequency stability dynamics.

The main difference between linear regression and the Tobit model is that in the Tobit model, instead of neglecting the information in the truncated or censored part of the data, the cumulative distribution function of that part of the data is used to calculate the log-likelihood function. The Tobit model log-likelihood incorporates both the normal distribution for the uncensored data and the cumulative normal distribution for the censored data to calculate the log-likelihood function. As a result, the coefficients of the learned function would be different from a simple regression model.

\Cref{loadshedding} shows the estimated optimal $p^{\text{\tiny UFLS}}$ after the outage of the generation unit $\ell$ at time $t$. Parameter $a^{\text{\tiny UFLS}}$ is the \gls{rocof} threshold above which \gls{ufls} potentially takes place. For disturbances with initial \glspl{rocof} lower than $a^{\text{\tiny UFLS}}$, no \gls{ufls} is needed. On the contrary, disturbances that produce larger \glspl{rocof} require \gls{ufls} to stabilize the system. In these cases, the optimal \gls{ufls} linearly varies with the initial \gls{rocof} and slope $b^{\text{\tiny UFLS}}$. 
\begin{equation} \label{loadshedding}
\begin{aligned}
    p_{\ell,t}^{\text{\tiny UFLS}} = \begin{cases}
    0 & \text{if $\Delta \dot f_{\text{\tiny initial}}\leq a^{\text{\tiny UFLS}}$} \\
    b^{\text{\tiny UFLS}} (\Delta \dot f_{\text{\tiny initial}} - a^{\text{\tiny UFLS}}) & \text{otherwise}
    \end{cases} & \text{\footnotesize $\;\;\;\;\forall t,\; \ell$}
\end{aligned}
\end{equation}

The obtained parameters for the Tobit model after being trained with the clustered dataset are as follows. The threshold for non-zero optimal \gls{ufls} estimations, $a^{\text{\tiny UFLS}}$, is equal to 1.66, and the slope $b^{\text{\tiny UFLS}}$ is equal to 3.246. The resulting model is illustrated with a dashed line in \Cref{fig:dataset}.

The next step is to linearize the conditional statement of \cref{loadshedding} by introducing the auxiliary binary variable $z_{\ell,t}$. The set of equations \cref{linearuflsrocof} shows the linearized version of \cref{loadshedding}. $M_2$ could be equal to the largest amount of optimal \gls{ufls} observed in the underlying training data.

\begin{subequations} \label{linearuflsrocof}
    \begin{align} 
    \Delta \dot f_{\text{\tiny initial}} &\leq a^{\text{\tiny UFLS}} + \frac{M_2}{b^{\text{\tiny UFLS}}} z_{\ell,t} &&\text{\footnotesize $\forall t,\; \ell$}\\
     p_{\ell,t}^{\text{\tiny UFLS}} &\leq M_2 z_{\ell,t} &&\text{\footnotesize $\forall t,\; \ell$}\\
     p_{\ell,t}^{\text{\tiny UFLS}} &\leq b^{\text{\tiny UFLS}} (\Delta \dot f_{\text{\tiny initial}} - a^{\text{\tiny UFLS}})+ M_2 (1 - z_{\ell,t}) &&\text{\footnotesize $\forall t,\; \ell$}\\
     p_{\ell,t}^{\text{\tiny UFLS}} &\geq b^{\text{\tiny UFLS}} (\Delta \dot f_{\text{\tiny initial}}- a^{\text{\tiny UFLS}})-M_2 ( 1 - z_{\ell,t}) &&\text{\footnotesize $\forall t,\; \ell$}
    \end{align}  
\end{subequations}

\section{Corrective FCUC formulation}\label{moduc}
The mathematical formulation of a standard \gls{uc} model is commonly presented in the form of a \gls{milp} as presented in the set of equations \Cref{baseuc}.
\begin{subequations}\label{baseuc}
\begin{align}
    \min_{u,p}\ c^{\text{g}}(p)&+c^{\text{suc}}(u)\label{of}
    \\
    u_{i,t}-u_{i,t-1}&=v_{i,t}-w_{i,t} &&\text{\footnotesize $\forall i,\; \forall t$}\label{bin1} 
    \\
    v_{i,t}+w_{i,t}&\leq1 &&\text{\footnotesize $\forall i,\;\forall t$}\label{bin2} 
    \\
    \sum_{s=t-\text{UT}_i+1}^{t}v_{i,s}&\leq u_{i,t} &&\text{\footnotesize $t\in\{\text{UT}_i,\dots, \mathcal{T}\}$}\label{ut} 
    \\
   \sum_{s=t-\text{DT}_i+1}^{t}w_{i,s}&\leq 1-u_{i,t} &&\text{\footnotesize $t\in\{\text{DT}_i,\dots, \mathcal{T}\}$}\label{dt} 
   \\
   p_{i,t}&\geq P^{\text{\tiny{min}}}_i u_{i,t} &&\text{\footnotesize $\forall i,\; \forall t$}\label{pmin}  
   \\
    p_{i,t}+r_{i,t}&\leq P^{\text{\tiny{max}}}_i u_{i,t} &&\text{\footnotesize $\forall i,\; \forall t$}\label{pmax} 
    \\
    p_{i,t}-p_{i,t-1}&\leq R^{\text{\tiny{up}}}_i &&\text{\footnotesize $\forall i,\; \forall t$}\label{ru} 
    \\
   p_{i,t-1}-p_{i,t}&\leq R^{\text{\tiny{down}}}_i &&\text{\footnotesize $\forall i,\; \forall t$}\label{rd} 
   \\
   \bigg(\sum_{i}p_{i,t}\bigg) +g^w_t+g^s_t& = \mathcal{D}_t &&\text{\footnotesize $\forall t$}\label{pb}
   \\
    {H}_\ell&\geq \frac{p_\ell}{2\Delta\dot f_{\text{crit}}}&&\text{\footnotesize $\forall t,\; \ell$}\label{rocof}
    \\
   \sum\limits_{i\neq \ell}r_{i,t}&\geq p_{\ell,t}&&\text{\footnotesize $\forall t,\; \ell$}\label{res}
\end{align}
\end{subequations}

The objective function is presented in \cref{of}. $c^{\text{g}}$ is the generation cost function in \euro, and $p$ is the vector of all online units' output power. Similarly, $c^{\text{suc}}$ is the start-up cost function in \euro, and $u$ is the binary variable indicating the commitment status of units.
\Cref{bin1,bin2} are the binary logic of the \gls{uc} formulation where $v$ and $w$ are the start-up and shut-down binary variables, respectively. The minimum up-time and down-time constraints are presented in \cref{ut,dt}. UT and DT are the minimum for the up-time and down-time of the generating units. The generators' output is limited by the minimum and maximum capacity and ramp rate bounds as shown in \cref{pmin,pmax,ru,rd}. $P^\text{\tiny{min}}$ is the minimum generation, and $P^\text{\tiny{max}}$ is the maximum generation capacity of the units. In addition, the reserve provided by each unit is represented by $r$. $R^\text{\tiny{up}}$ and $R^\text{\tiny{down}}$ are the maximum ramp-up and ramp-down of generating units, respectively.
\Cref{pb} represents the power balance. The wind and solar generation are represented by $g^w$ and $g^s$ based on historical data. $\mathcal{D}_t$ is the total demand in each time interval.
In small power systems where the total inertia is low, the \gls{rocof} constraint presented in \cref{rocof} is frequently used to ensure that the \gls{rocof} wouldn't exceed a critical value by ensuring the availability of sufficient inertia. Here, $H_{\ell}$ is the summation of inertia constants of online generating units, $\Delta \dot f_{\text{crit}}$ is the critical \gls{rocof}, and $p_\ell$ is power lost corresponding to the outage of unit $\ell$.
The reserve constraint is presented in \cref{res} that makes sure there is enough headroom available to compensate for the outage of generating unit $\ell$.

Now the basic \gls{uc} formulation needs to be modified in order to add the optimal \gls{ufls} estimation into the formulation. First, the initial \gls{rocof} has to be presented as a decision variable. \Cref{defrocof2} computes the absolute value of the initial \gls{rocof} variable after an active-power disturbance $p_{\ell,t}$ described by the aforementioned loss of generation unit $\ell$. 
\begin{equation}
    2\Delta \dot f_{\text{\tiny initial}} \sum_{i\neq \ell}\big(u_{i,t} H_i \big) = p_{\ell,t} \;\;\;\text{\footnotesize $\forall t,\; \ell$} \label{defrocof2}
\end{equation}

The bilinear product in \cref{defrocof2} can be linearized by defining an auxiliary continuous variable, $y_{i,\ell,t}$ that represents the product of $\Delta\dot f_{\text{\tiny initial}}$ and $u_{i,t}$ after an outage of unit $\ell$, and by applying the big-M method. The linear equivalent formulation is shown in the set of equations \cref{linrocof}. $M_1$ could be equal to $\Delta f_{crit}$.

\begin{subequations}\label{linrocof}
    \begin{align}
        2\sum_{i\neq \ell} &y_{i,\ell,t}H_i = p_{\ell,t}&&\text{\footnotesize $\forall t,\; \ell$}\\
        y_{i,\ell,t} &\leq M_1 u_{i,t}&&\text{\footnotesize $\forall t,\; \ell$}\\
        y_{i,\ell,t} &\leq \Delta \dot f_{\text{\tiny initial}}&&\text{\footnotesize $\forall t,\; \ell$}\\
        y_{i,\ell,t} &\geq \Delta \dot f_{\text{\tiny initial}} - M_1(1-u_{i,t})&&\text{\footnotesize $\forall t,\; \ell$}
    \end{align}
\end{subequations}

The static reserve criterion in \cref{res} is then relaxed and becomes dynamic by considering an estimated amount of optimal \gls{ufls} as shown in \cref{newres}. Note that for a sufficiently large outage in \glspl{ips}, the \gls{ufls} scheme is activated no matter how much reserve is provided. According to that, it is argued that the reserve requirement in \cref{res} is excessive, leading to overly conservative solutions. Therefore, \cref{newres} is presented here with the additional term $p^{\text{\tiny UFLS}}_\ell$, which is the estimated optimal \gls{ufls} explained in the previous section.
\begin{align}
    \sum_{ \substack{ i\neq \ell}} r_{i,t} &\geq p_{\ell,t} - p_{\ell,t}^{\text{\tiny UFLS}} &&\text{\footnotesize $\forall t,\; \ell$}
\label{newres}
\end{align}

The objective function in \cref{of} should also be modified to penalize \gls{ufls} to avoid solutions that increase the optimal \gls{ufls}. 
\begin{equation}
    \min\limits_{u, p} c^g(p)+c^{\text{suc}}(u)+c^{\text{\tiny UFLS}}(p^{\text{\tiny UFLS}}) \label{objnew}
\end{equation}

To effectively manage the contrasting impact of cost savings and the risk of load-shedding, \gls{ufls} is optimized alongside the operation costs. The cost of \gls{ufls} represented by $c^{\text{\tiny UFLS}}(.)$ must be determined by multiplying the post-outage cost of \gls{ufls} by the probability of the outage occurring to quantify the risk of \gls{ufls} \cite{o2021probabilistic}:
\begin{equation}
    c^{\text{\tiny UFLS}}(p^{\text{\tiny UFLS}})=\sum_t\sum_i C^o\times\text{FOR}_i\times p^{\text{\tiny UFLS}}_{\ell,t}\label{uflscost}
\end{equation}
where $C^o$ is the post-outage cost of \gls{ufls} in \euro/MW and FOR is the forced outage rate of generator $i$.
Note that throughout the formulations, the indices $t$, $i$, $\ell$, and $s$ are used for the time, generating units, disturbance, and time alias sets, respectively.
The resulting corrective \gls{fcuc} formulation includes the objective function in \cref{objnew} 
subject to \cref{uflscost,bin1,bin2,ut,dt,pmax,pmin,ru,rd,pb,rocof,linrocof,newres,linearuflsrocof}.

\section{Corrective FCUC results}\label{sec:result}
\subsection{Case study}\label{case}
The proposed formulation is implemented in a real Spanish \gls{ips}.
There are eleven diesel generators in the system, and about 10\% of its annual demand is provided by wind and solar power generation. This power system has a peak demand of around 40 MW. Inertia and spinning reserve are currently provided by conventional generation units, and virtual inertia potentially provided by \glspl{res} is not considered. \Cref{tab:genparam} shows the technical parameters of the generators.
\begin{table}[t!]
    \centering
    \caption{Parameters of the Generating Units}
    \begin{adjustbox}{width=0.7\textwidth}
    \begin{tabular}
    {c ccccc}
    \toprule
        Unit & $P^{\text{max}}$ (MW) & $P^{\text{min}}$ (MW)  & $M^{\text{base}}$ (MVA) & $H$ (s) & $K$ (pu) \\ \midrule
        1 & 3.82 & 2.35 &  5.4 & 1.75 & 20\\
        2 & 3.82 & 2.35 &  5.4 & 1.75 & 20\\
        3 & 3.82 & 2.35 &  5.4 & 1.75 & 20\\
        4 & 4.3 & 2.82 &  6.3 & 1.73 & 20\\
        5 & 6.7 & 3.3 & 9.4 & 2.16 & 20\\
        6 & 6.7 & 3.3 & 9.6 & 1.88 & 20 \\
        7 & 11.2 & 6.63 & 15.75 & 2.1 & 20 \\
        8 & 11.5 & 6.63 & 14.5 & 2.1 & 20\\
        9 & 11.5 & 6.63 & 14.5 & 2.1 & 20\\
        10 & 11.5 & 6.63 &  14.5 & 2.1 & 20\\ 
        11 & 21 & 4.85 &  26.82 & 6.5 & 21.25\\ \bottomrule
    \end{tabular}
    \end{adjustbox}
    \label{tab:genparam}
\end{table}

\subsection{Sample day analysis}\label{moducres}
To evaluate the performance of the proposed corrective \gls{fcuc} formulation, simulations are performed for a sample day in three different cases.
\begin{itemize}
    \item Case I: the standard \gls{uc} formulation.
    \item Case II: the corrective \gls{fcuc} including the data-driven estimation of optimal \gls{ufls} without penalizing \gls{ufls} in the objective function.
    \item Case III: the corrective \gls{fcuc} including the data-driven estimation of optimal \gls{ufls} with post-outage cost of \gls{ufls} 100 k\euro/MW.
\end{itemize}
Note that the actual cost of \gls{ufls} may vary depending on the specific system and the desired level of conservativeness. \Cref{sensitivity} provides a sensitivity analysis on the effects of considering different post-outage costs for \gls{ufls}. Additionally, an outage rate of 2 occurrences per year is assumed.
\Cref{tab:results} shows the operation costs, the total \gls{ufls} costs, the amount of estimated optimal \gls{ufls} using the Tobit model, and the amount of optimal \gls{ufls} calculated using detailed simulations for different cases.
\begin{table}[t!]
    \centering
    \caption{Costs and \gls{ufls} amounts of different cases}
    \begin{adjustbox}{width=1\textwidth}
    \begin{tabular}
        {c cccc}
        \toprule
       Case & Operation cost (\euro) & \gls{ufls} cost (\euro) & \thead{Estimated optimal \\ \gls{ufls} (MW)} & \thead{Calculated optimal \\ \gls{ufls} (MW)} \\\midrule
        Case I & 70207.20 & - & - & - \\
        Case II & 68951.70 & 0 & 106.09 & 118.87 \\
        Case III & 69778.30 & 738.60 & 32.35 & 42.09 \\
        \bottomrule
    \end{tabular}
    \end{adjustbox}
    \label{tab:results}
\end{table}
 According to \Cref{tab:results}, the corrective \gls{fcuc} successfully lowers the operation costs by utilizing the estimation of optimal \gls{ufls} and relaxing the reserve constraints. The reduction of operation costs is higher in Case II compared to Case III. This indicates that in Case II, the formulation results in higher \gls{ufls} amounts to reduce the operation cost as much as possible. However, the trade-off between cost and security is balanced in Case III by adding \gls{ufls} cost to the objective function. The optimal \gls{ufls} calculated using detailed simulations is shown here to demonstrate the accuracy of the estimation. It is important to note that the estimation is lower than the exact optimal amount of \gls{ufls}, indicating that the reserve constraint is relaxed on the conservative side, so the security of the system is not jeopardized at any point. 
 
\begin{figure}[!htbp]
\centering
\begin{subfigure}{0.9\linewidth}
    \includegraphics[width=\linewidth]{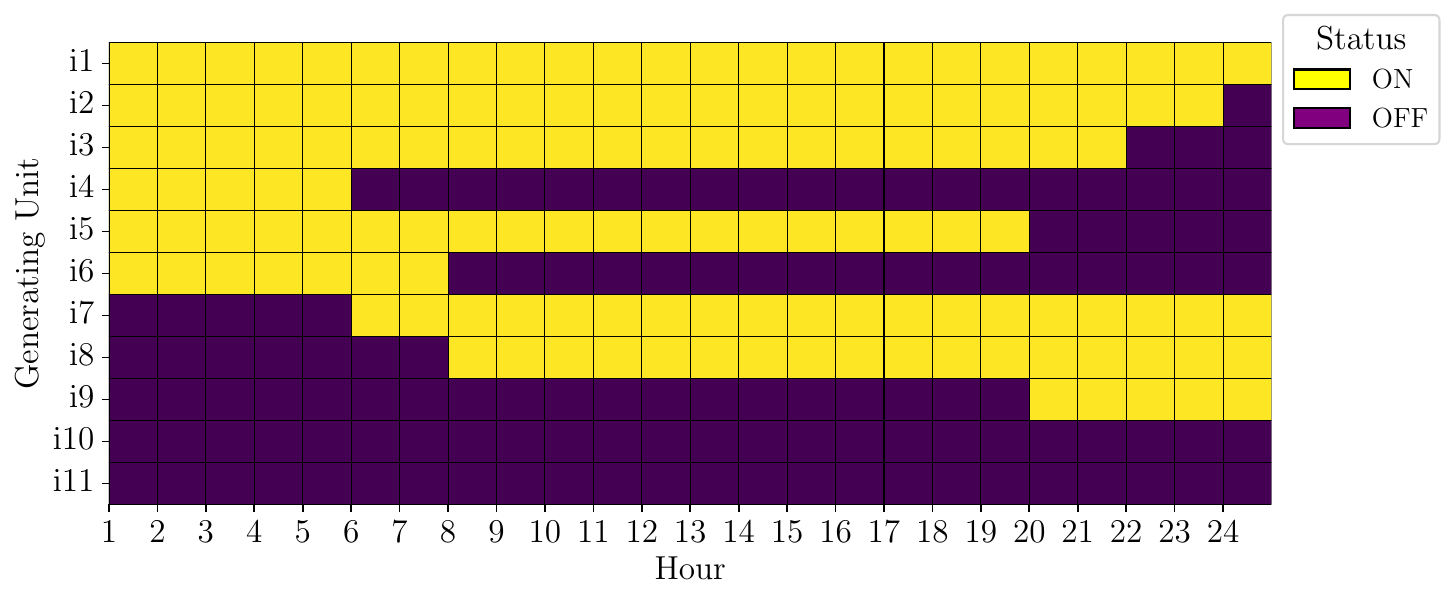}
    \caption{\scriptsize On/Off status of generating units Case I}\label{Fig:on-UC}
\end{subfigure}
\begin{subfigure}{0.9\linewidth}
    \includegraphics[width=\linewidth]{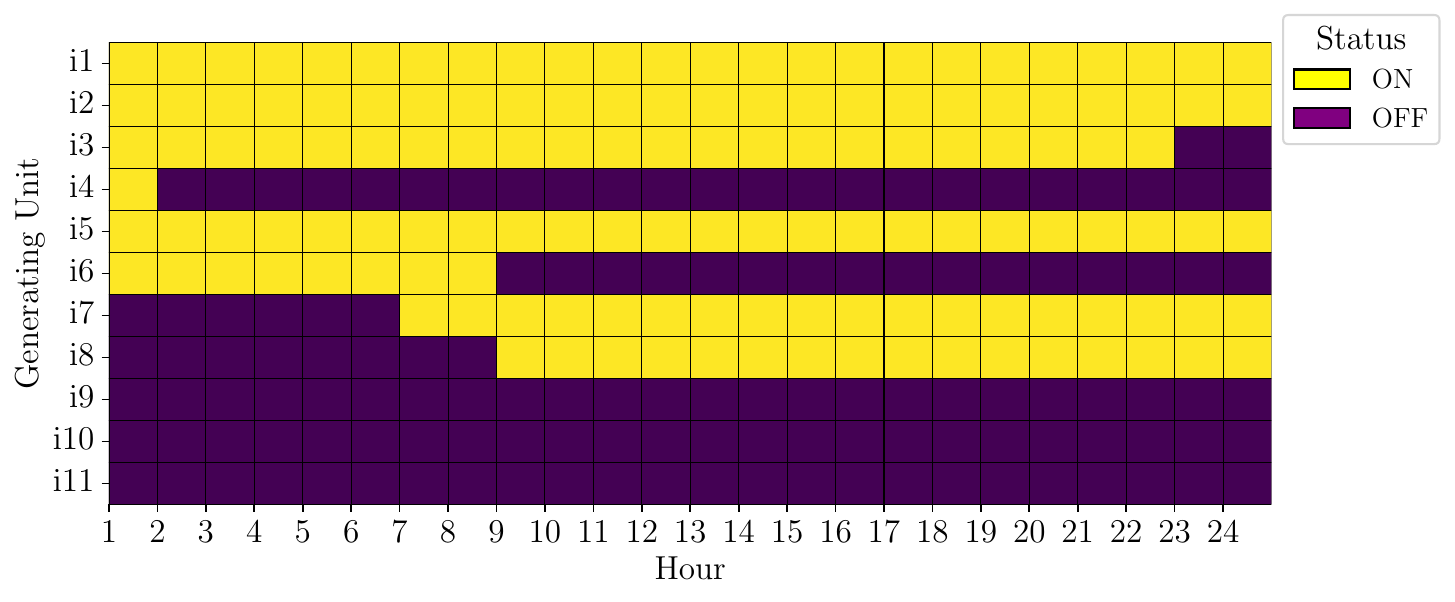}
    \caption{\scriptsize On/Off status of generating units Case II}\label{Fig:on-tobit0}
\end{subfigure}
\begin{subfigure}{0.9\linewidth}
    \includegraphics[width=\linewidth]{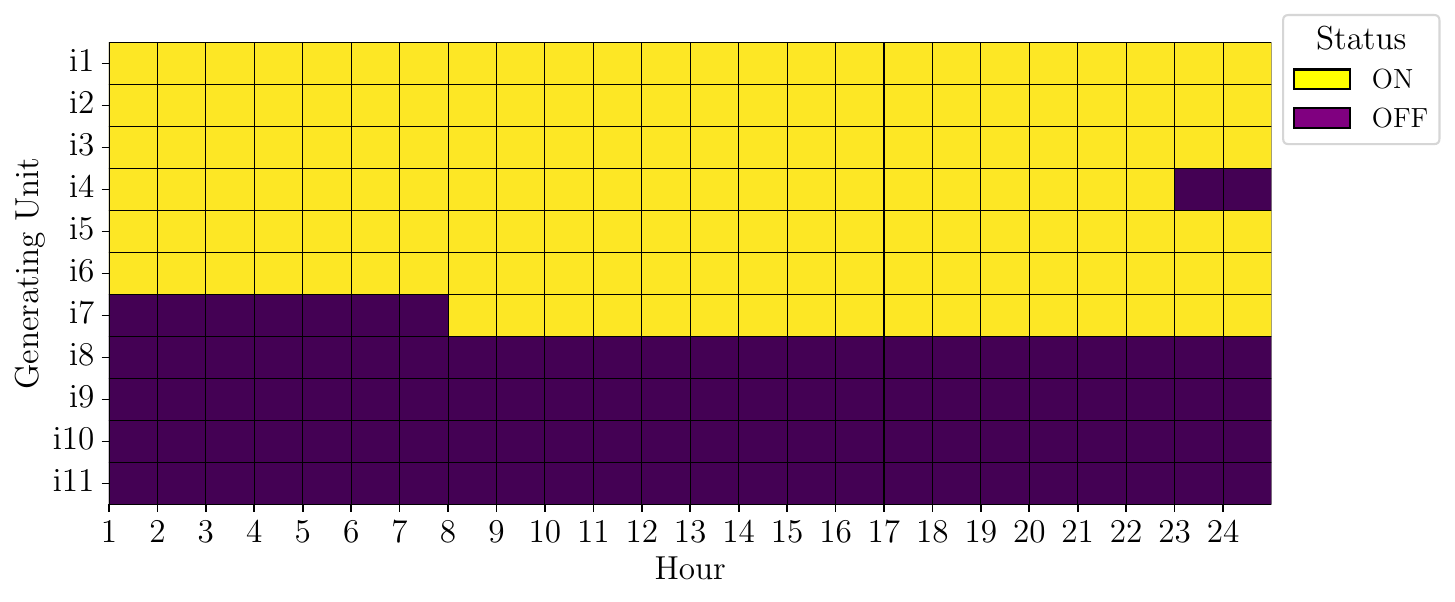}
    \caption{\scriptsize On/Off status of generating units Case III} \label{Fig:on-tobit100}
\end{subfigure}
\caption{Day-ahead commitment decisions}\label{fig:on}
\end{figure}

\Cref{fig:on} illustrates the on/off status of generators for the three cases. It is observed in \Cref{fig:on} that the commitment of generators in Case I is different from those in other cases. For example, in Case I, unit $i_4$ is turned off after time step 6, but in Case II, this unit is turned off after time step 1. Also, unit $i_9$ is not turned on at all in Case II. This indicates that in Case I, a higher price is paid to supply the reserve requirement. In Case III, units $i_4$ and $i_6$ keep generating, and units $i_8$ and $i_9$ are not committed at all. These results demonstrate the effectiveness of the proposed corrective \gls{fcuc} in reducing the operation costs when benefiting from the optimal \gls{ufls} estimation.

\begin{figure}[t!]
    \centering
    \begin{subfigure}{1\linewidth}
        \includegraphics[width=\linewidth]{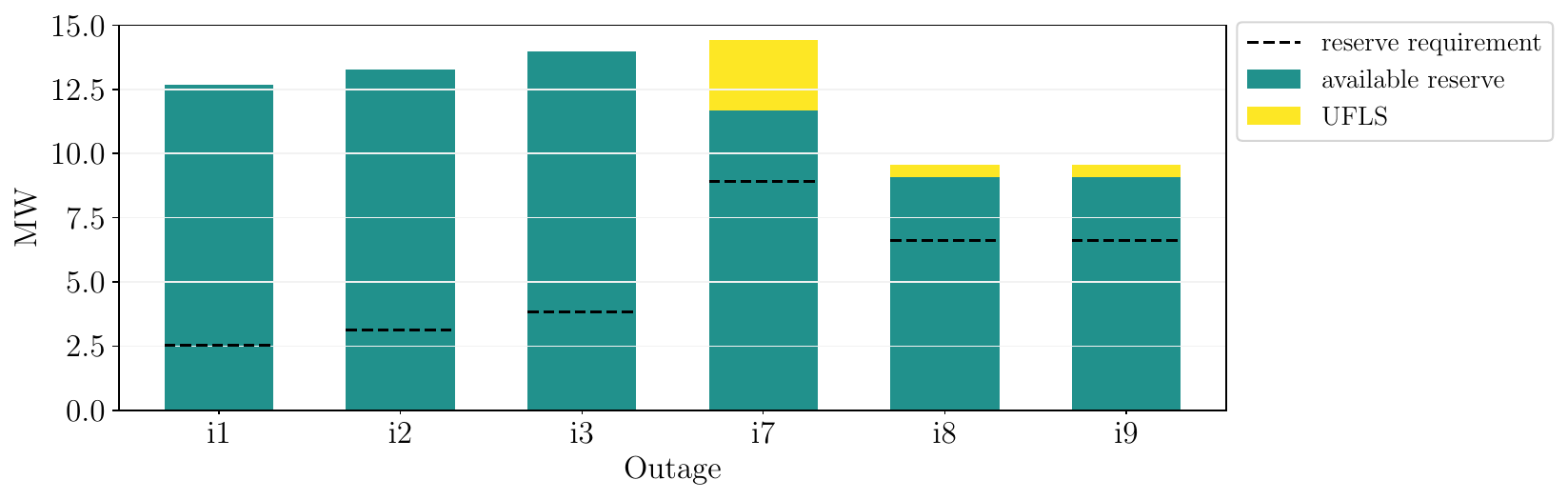}
        \caption{Case I}
        \label{fig:oneuc} 
    \end{subfigure}
    \begin{subfigure}{1\linewidth}
        \includegraphics[width=\linewidth]{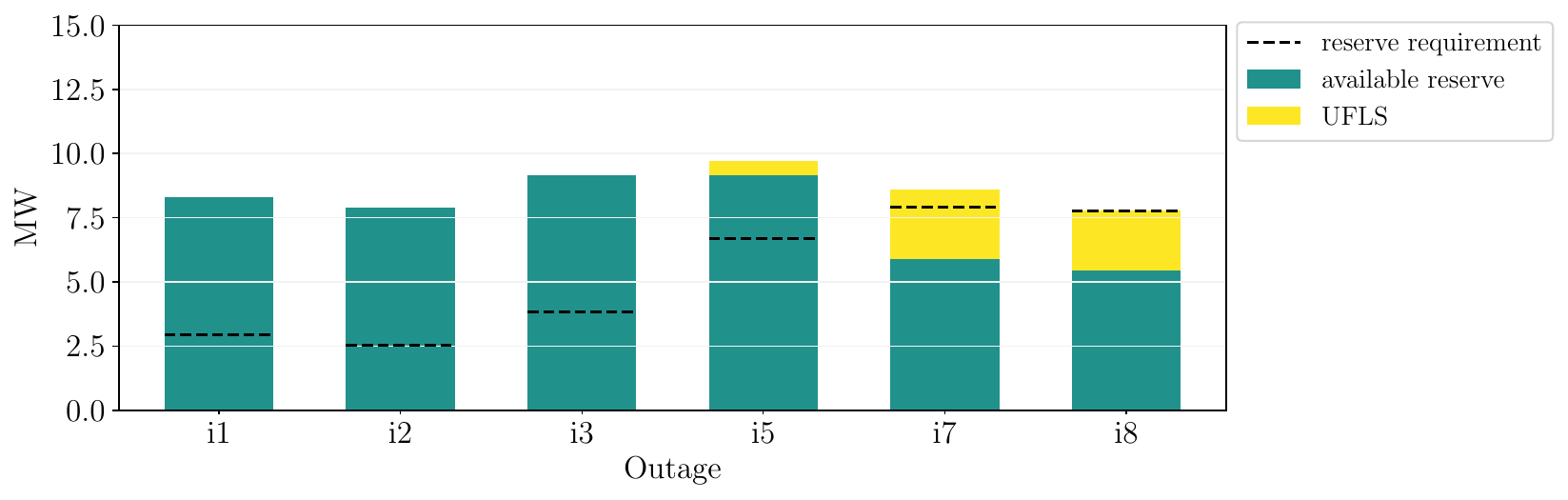}
        \caption{Case II}
        \label{fig:onetobit0} 
    \end{subfigure}
    \begin{subfigure}{1\linewidth}
        \includegraphics[width=\linewidth]{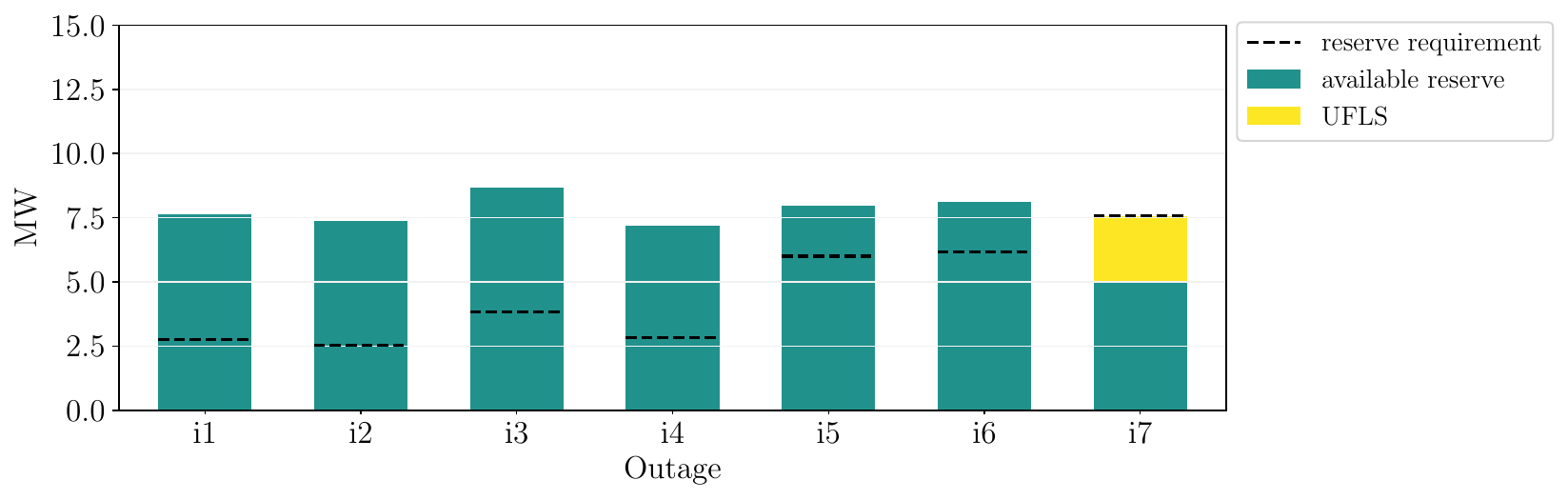}
        \caption{Case III}
        \label{fig:onetobit100}
    \end{subfigure}
\caption{Reserve and UFLS estimation for each possible outage in time step 21}\label{fig:onehour}
\end{figure}

To provide additional insight into the basis of cost reductions, \Cref{fig:onehour} illustrates the amount of available reserve, the amount of reserve requirement, and the amount of estimated \gls{ufls} for every possible outage in time step 21 for all cases.
For Case I illustrated in \Cref{fig:oneuc}, although the outage of units $i_7$, $i_8$, and $i_9$ leads to positive amounts of optimal \gls{ufls}, the formulation does not allow their usage of these amounts to relax the reserve requirement.
In Case II shown in \Cref{fig:onetobit0}, it is observed that for the outage of units $i_5$, $i_7$, and $i_8$, the estimated optimal \gls{ufls} is positive. In addition, for the outage of units $i_7$ and $i_8$, the estimated \gls{ufls} amounts are utilized as part of the reserve requirement. However, in Case III illustrated in \Cref{fig:onetobit100}, where \gls{ufls} is penalized in the objective function, the generating units are different and only the outage of unit $i_7$ leads to \gls{ufls}, which is then used as part of the reserve requirement. This indicates that the level of conservativeness can be balanced by changing the \gls{ufls} cost parameter.

\subsection{Analysis of different daily load profiles}\label{boxplot}
\begin{figure}[t!]
    \centering
    \includegraphics[width=1\linewidth]{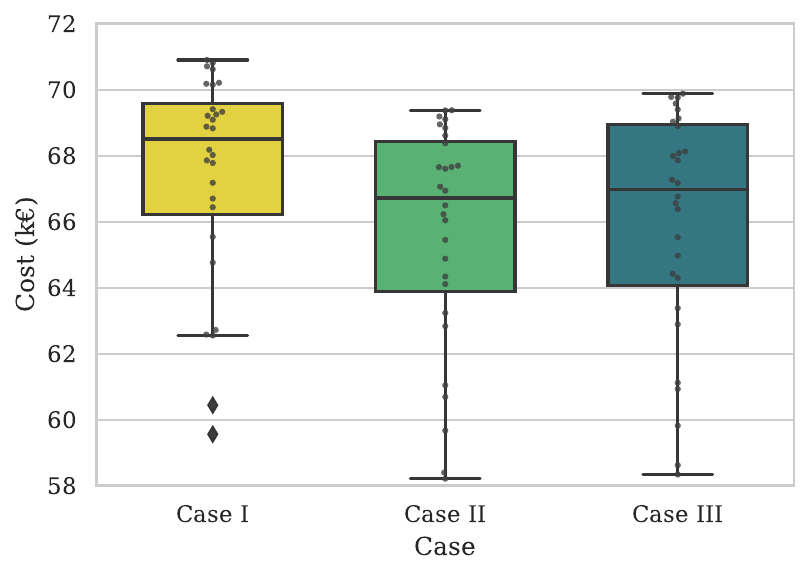}
    \caption{Operation costs of various days for each case}
    \label{fig:boxplot}
\end{figure}

The scheduling problem is solved for every case using the data of one sample week in each season in order to test the effectiveness of the proposed corrective \gls{fcuc} formulation regarding different daily load profiles. \Cref{fig:boxplot} shows the resulting operation costs of 28 days for the three cases.  

It is shown that the operation cost in Case I is constantly higher than both Case II and Case III. Moreover, when comparing Case II and Case III, it is observed that the operation costs in Case II are consistently lower than those in Case III due to no \gls{ufls} cost in the objective function in Case II.

\subsection{Sensitivity analysis of optimal UFLS cost}\label{sensitivity}
\begin{figure}[t!]
    \centering
    \includegraphics[width=1\linewidth]{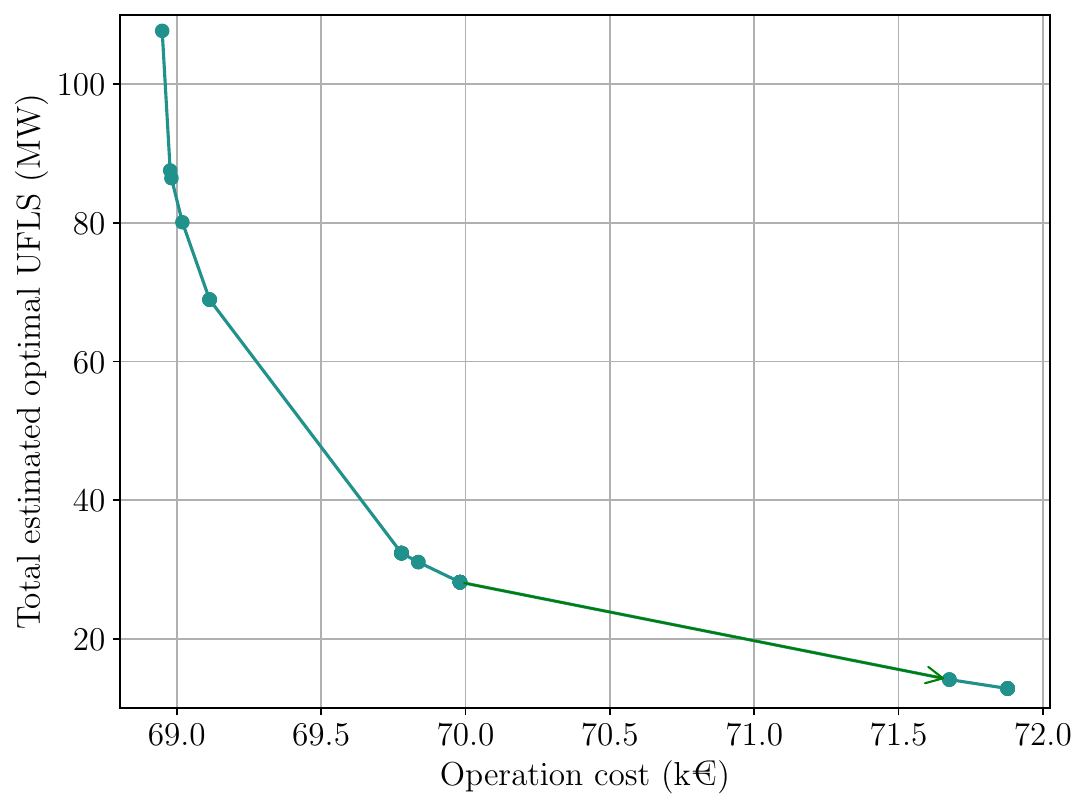}
    \caption{Sensitivity analysis of the cost associated with potential UFLS}
    \label{fig:pareto}
\end{figure}

This section presents a sensitivity analysis for the cost associated with the estimated optimal amounts of \gls{ufls} using the Tobit model. For this purpose, the post-outage cost of \gls{ufls} in \euro/MW shown in \cref{uflscost} has been varied from 0 to 1 million euros with increments of 10 k\euro. Note that the value of load shedding can arise from a cost the regulator imposes for energy not served or from demand bids for interruption. \Cref{fig:pareto} shows and compares the system operation costs and the estimated amount of optimal \gls{ufls}. The arrows show the increase of \gls{ufls} cost. As it can be inferred, larger \gls{ufls} penalty factors reduce the use of optimal \gls{ufls}, and favor a preventive dispatch of generating units that provide more inertia and primary frequency control. This leads to higher system operation costs. Conversely, smaller \gls{ufls} penalty factors favor more \gls{ufls}. Therefore, the penalty factor has a strong impact. It can further be used as a parameter in the model that is available to the system operator in order to modify the level of conservativeness while balancing the trade-off between costs and security. Note that in total, one hundred different values for the post-outage cost of \gls{ufls} are analyzed. However, a lot of these values give the same results, so the number of points visible in \Cref{fig:pareto} is less than that. 

\subsection{Compatibility with the conventional UFLS scheme}\label{optimalvsreal}
The framework presented in this paper enables estimating optimal \gls{ufls} in order to be utilized inside the scheduling problem. The \gls{ufls} scheme here is assumed to be based on the optimal amounts. The estimation here is on the more conservative side to make sure the security of the system is not compromised by relaxing the reserve constraint too much. 

\begin{figure}[t!]
    \centering
    \begin{subfigure}{0.49\linewidth}
        \includegraphics[width=\linewidth]{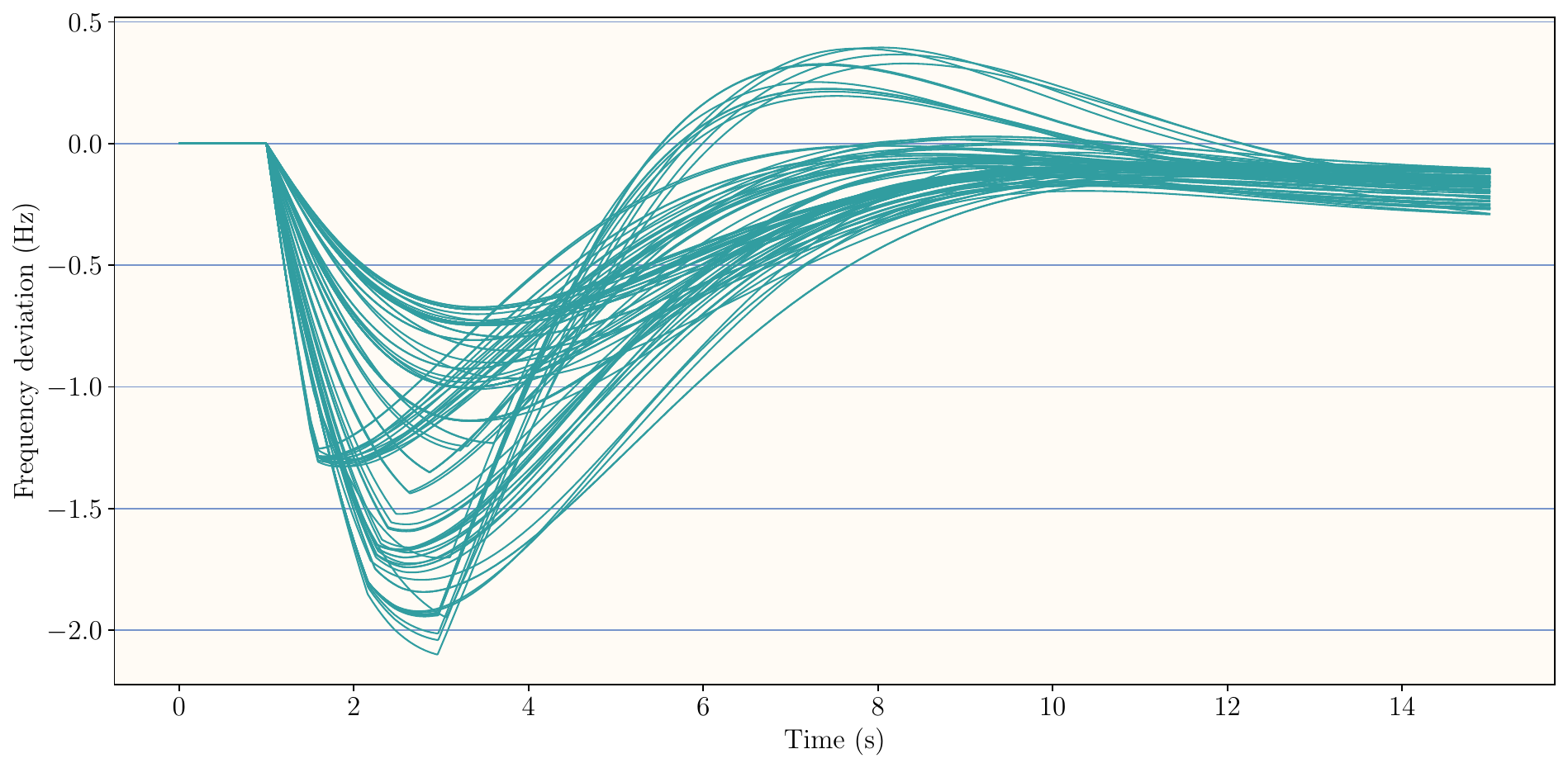}
        \caption{standard UC}
        \label{fig:frequc} 
    \end{subfigure}
    \begin{subfigure}{0.49\linewidth}
        \includegraphics[width=\linewidth]{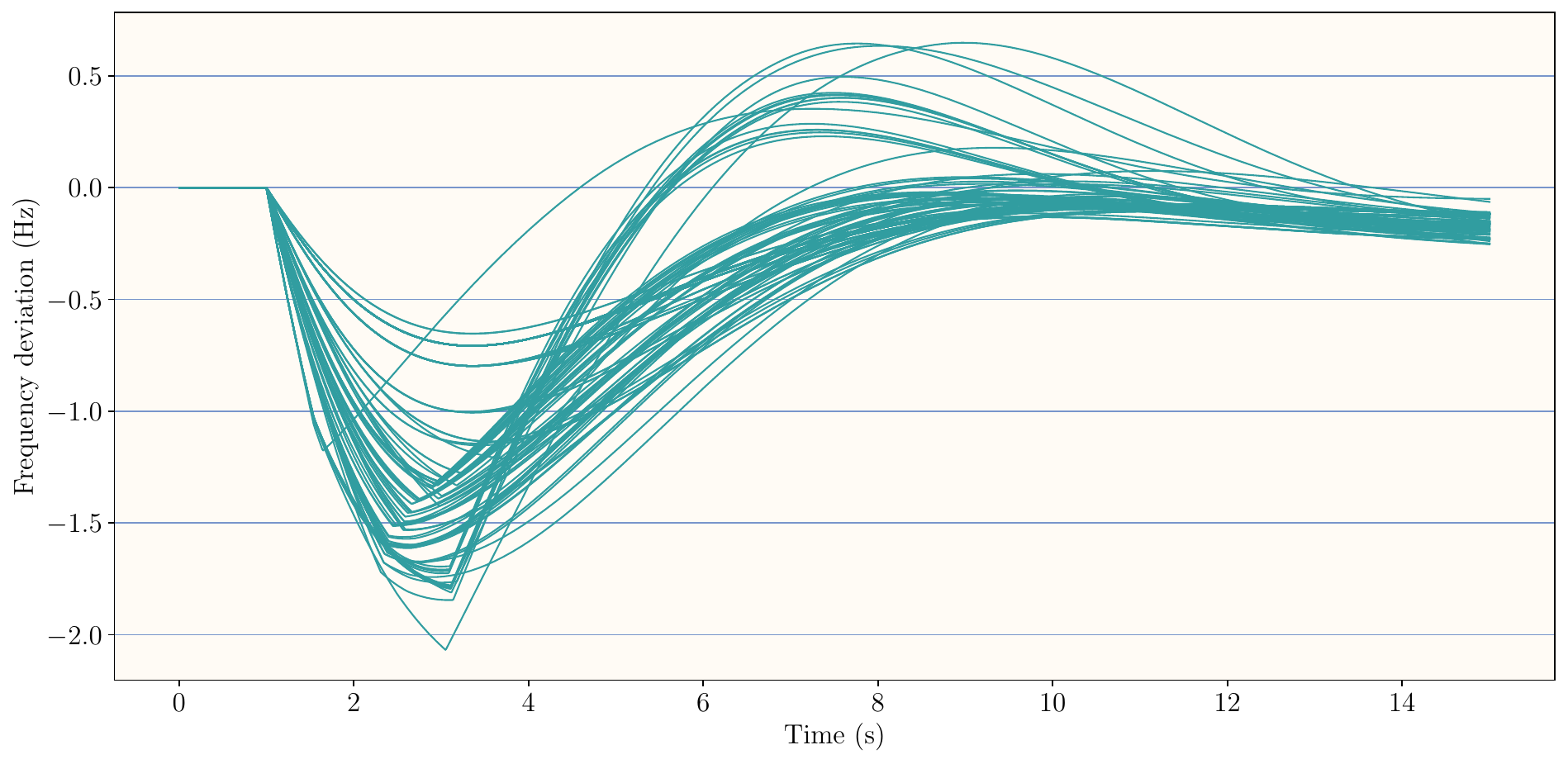}
        \caption{corrective FCUC}
        \label{fig:freqtobit}
    \end{subfigure}
\caption{Post-contingency frequency performance of different cases}\label{fig:freq}
\end{figure}

This framework is compatible with conventional \gls{ufls} as well. The conventional schemes operate in a step-wise manner with predefined activation thresholds and time delays. 
Detailed dynamic simulations are performed to analyze the effectiveness of the proposed formulations, considering the conventional scheme that is currently implemented in the real Spanish \gls{ips}, which is the case study of this paper.
\Cref{fig:freq} shows the post-contingency frequency performance for standard \gls{uc} and the proposed corrective \gls{fcuc} formulation, for every possible outage in a sample day. This further illustrates the preserved security of the system after relaxing the reserve requirement constraint. Comparing \Cref{fig:frequc} with \Cref{fig:freqtobit}, it is observed that in Case II and Case III, the frequency stability of the system is not compromised. Additionally, it is seen that the occurrences of frequency falling below 48 Hz are even lower in the proposed formulation compared to the standard \gls{uc}.

\section{Conclusions}\label{sec:conclusion}
This paper has presented a novel corrective \gls{fcuc} for \glspl{ips} implementing data-driven constraint learning techniques to estimate the amount of optimal \gls{ufls}. The proposed formulation enables co-optimization of generation dispatch and \gls{ufls}. In other words, the power system operates at minimum cost but is still secure after the outage of each generation unit. The proposed formulation modifies the currently used static spinning reserve criterion by accounting for optimal \gls{ufls} amounts, which are estimated as a function of the initial post-disturbance \gls{rocof} using the Tobit model. 

The data-driven corrective \gls{fcuc} formulation has been applied to a Spanish \gls{ips} with a peak demand of 40 MW. The performance of the proposed formulation in terms of system operation cost and amount of optimal \gls{ufls} has been compared with the performance of a standard \gls{uc}. It is shown that the corrective \gls{fcuc} can reduce system operation costs without necessarily increasing \gls{ufls}. 
The performance of the formulation depends on the parameter that represents the cost associated with the optimal \gls{ufls} readily available to modify and adjust the level of conservativeness desired. A sensitivity analysis has been performed to show the effects of this parameter in the modified formulation.
Time-domain simulations have shown that the proposed framework is also compatible with conventional \gls{ufls} schemes and does not deteriorate the security of the system when the conventional schemes are applied.


\section*{Acknowledgment}

This research has been funded by grant PID2022-141765OB-I00 funded by MCIN/AEI/ 10.13039/501100011033 and by “ERDF A way of making Europe”.

 \bibliographystyle{elsarticle-num} 
 \bibliography{cas-refs}
 
\end{document}